%% file: main.tex
\documentclass[conference]{IEEEtran}
\IEEEoverridecommandlockouts
\usepackage{cite}
\usepackage{amsmath,amssymb,amsfonts}
\usepackage{algorithmic}
\usepackage{graphicx}
\usepackage{textcomp}
\usepackage{xcolor}
\usepackage{url}

\usepackage[normalem]{ulem}

\newcommand{\ours}[1]{\textsc{KGACG}}

\def\BibTeX{{\rm B\kern-.05em{\sc i\kern-.025em b}\kern-.08em
    T\kern-.1667em\lower.7ex\hbox{E}\kern-.125emX}}
\begin{document}

\title{Knowledge-Guided Multi-Agent Framework for Application-Level Software Code Generation}
% \title{Feedback-Based Multi-Agent Framework for Application-Level Software Code Generation
% \\

\author{
\IEEEauthorblockN{Qian Xiong}
\IEEEauthorblockA{
\textit{School of Information Science \& Technology} \\
\textit{Beijing Forestry University}\\
Beijing, China \\
x\_qianq@bjfu.edu.cn}
\and
\IEEEauthorblockN{Bo Yang*}
\IEEEauthorblockA{\textit{School of Information Science \& Technology} \\
\textit{Beijing Forestry University}\\
Beijing, China \\
yangbo@bjfu.edu.cn}
\and
\IEEEauthorblockN{Weisong Sun*}
\IEEEauthorblockA{
\textit{NanyangTechnological University}\\
Singapore \\
weisong.sun@ntu.edu.sg}
\and
\IEEEauthorblockN{Yiran Zhang}
\IEEEauthorblockA{
\textit{Nanyang Technological University}\\
Singapore \\
yiran002@e.ntu.edu.sg}
\and
\IEEEauthorblockN{Tianlin Li}
\IEEEauthorblockA{
\textit{Nanyang Technological University}\\
Singapore \\
tianlin001@e.ntu.edu.sg}
\and
\IEEEauthorblockN{Yang Liu}
\IEEEauthorblockA{
\textit{Nanyang Technological University}\\
Singapore \\
yangliu@ntu.edu.sg}
\and
\IEEEauthorblockN{Zhi Jin}
\IEEEauthorblockA{
\textit{Wuhan University}\\
Wuhan,China\\
zhijin@whu.edu.cn}
}

\maketitle

\input{sections/abstract}


\input{sections/introduction}
% \input{sections/related work}
\input{sections/methodology}
\input{sections/case_study}

\input{sections/challenges_and_opportunities}
\input{sections/conclusion}

\bibliography{ASE.bib}
\bibliographystyle{IEEEtran}

\end{document}

%% file: sections/abstract.tex
\begin{abstract}
Automated code generation driven by Large Language Models (LLMs) has enhanced development efficiency, yet generating complex application-level software code remains challenging. Multi-agent frameworks show potential, but existing methods perform inadequately in large-scale application-level software code generation, failing to ensure reasonable organizational structures of project code and making it difficult to maintain the code generation process. 
To address this, this paper envisions a Knowledge-Guided
Application-Level Code Generation framework named \ours{}, which aims to transform software requirements specification and architectural design document into executable code through a collaborative closed-loop of the Code Organization \& Planning Agent (COPA), Coding Agent (CA), and Testing Agent (TA), combined with a feedback mechanism. We demonstrate the collaborative process of the agents in \ours{} in a Java Tank Battle game case study while facing challenges. \ours{} is dedicated to advancing the automation of application-level software development.
\end{abstract}

%% file: sections/introduction.tex
\begin{figure*}[hbtp]
    \centering
    \includegraphics[width=0.78\textwidth]{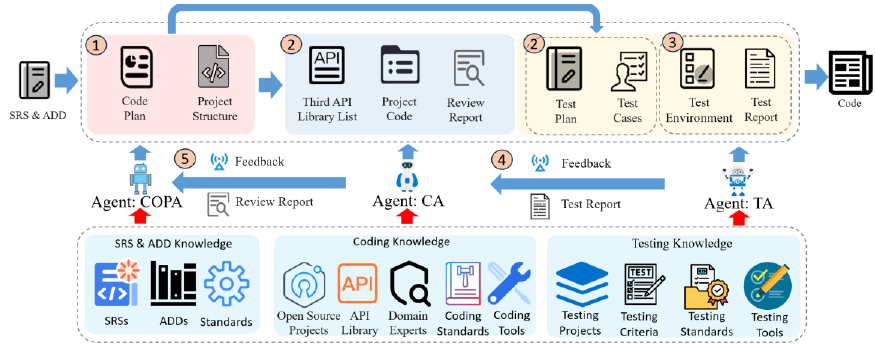}
    \caption{The Overview of Knowledge-Guided Multi-Agent Framework for Application-Level Software Code Generation}
    \label{Fig: overview}
\end{figure*}

\section{Introduction}
Automated code generation~\cite{liu2025sew, ni2024tree, li2025codetree, zhang2023self, mu2023clarifygptempoweringllmbasedcode} driven by LLMs has rapidly moved from synthesising short snippets to assisting across the full software development lifecycle \cite{jiang2024survey, fakhoury2024llm, liu2024exploring, nejjar2025llms, liao20243, zhang2024codeagent, chen2021evaluating, zheng2023codegeex}.
Industrial tools such as GitHub Copilot \cite{chen2021evaluating} and open source systems such as CodeGeeX \cite{zheng2023codegeex} already improve developer productivity on routine coding tasks.
Nevertheless, when the target shifts to complex, application-level programs, single LLM pipelines reveal three systematic deficits:
(1) context isolation: the model sees only the current prompt and cannot reason over multi-file repositories \cite{jiang2024self, le2023codechain};
(2) lack of grounding: generated code is rarely executed or tested, allowing silent semantic errors to propagate \cite{dong2024self, pan2025codecor};
(3) absence of iteration: the session ends after one response, preventing the self-correction loops that human developers rely on \cite{madaan2023self, chenteaching}.

Recent research has explored the use of multi-agent frameworks that decompose the code generation process into specialized stages, assigning distinct roles, such as document analysis, code synthesis, and testing, to dedicated agent \cite{qian2024chatdevcommunicativeagentssoftware, metagpt2024, ishibashi2024selforganizedagentsllmmultiagent, nunez2024autosafecodermultiagentframeworksecuring, huang2024agentcodermultiagentbasedcodegeneration, zhang2024pair, lin2025soen}.
Feedback loops (compiler messages, unit test verdicts, runtime traces) are continuously appended to a shared workspace, turning the LLM from a passive generator into an autonomous participant in an observe–plan–act cycle \cite{jiang2024self, le2023codechain, dong2024self}. Research on multi-agent code generation is still maturing, early results already show 20–40\% relative gains in correctness and maintainability compared with single-model baselines \cite{islam2024mapcodermultiagentcodegeneration, yuksel2024multiaiagentautonomousoptimization, zhu2025adacoderadaptiveplanningmultiagent}.

However, existing methods struggle with large-scale code projects: most perform well in small-scale tasks but fail to address industrial complexities (such as multicomponent coordination, diverse requirement adaptation, and long-term maintainability assurance). Specifically, generated code often neglects long-term maintenance needs, reducing readability and accumulating technical debt~\cite{perera2024systematic} in large-scale iterations. Bridging the gap between lab research and industrial application remains a critical challenge.

To fill this gap, we propose an automated code generation framework, \textit{Knowledge-Guided
Application-Level Code Generation (\ours{})}. The overview of \ours{} is shown in Figure~\ref{Fig: overview}. \ours{} enables collaboration among three architectural agents to generate code jointly, based on input Software Requirements Specification (SRS) and Architectural Design Document (ADD). To support these agents, we implement a feedback mechanism that incorporates message feedback and artifact maintenance to tackle key challenges in code generation.

% \vspace{-0.5em}

%% file: sections/methodology.tex
\section{\ours{} Framework}

\label{sec:our_framework}
% \ws{Keep the Methodology content within one and a half pages (i.e., three columns).}
Figure~\ref{Fig: overview} illustrates the architecture of \ours{}. Given a SRS and an ADD, \ours{} generates the corresponding source code through the synergistic collaboration of three specialized autonomous agents: COPA, CA, TA. The CA, responsible for code generation, incorporates feedback from the TA’s test reports to refine logic and optimize algorithms. Concurrently, it utilizes architectural updates from COPA to incrementally refactor conflicting modules. This feedback-driven optimization cycle promotes iterative collaboration among the three agents. Upon automatic compilation and error detection by the CA, a feedback loop is initiated. COPA cross-references these error logs with the original SRS and ADD to pinpoint root causes, adjust misunderstood constraints, revise the code implementation plan, and synchronize these updates with the CA. Subsequently, the CA employs these revised plans to incrementally rebuild the code, reusing non-conflicting modules to minimize redundancy, and initiates compilation validation. Simultaneously, the TA furnishes structured test reports to the CA to facilitate code enhancement. For persistent or complex issues, the process escalates to manual auditing, ensuring that iterative optimization aligns with both technical specifications and practical implementation requirements. To further enhance the capabilities of these three agents, external knowledge was strategically integrated into their workflows.  

\subsection{Code Organization \& Planning Agent (COPA)}
% COPA bridges detailed requirements/architecture documents and actual coding. Its core role is to convert SRS and ADD content into structured, actionable code implementation plans to guide subsequent coding. Inputs are detailed SRS and ADD; output is a specific code implementation plan for coding and a structured project framework. To achieve this, COPA initially performs a thorough analysis and extracts key information from the input documents, transforms such information into specific code implementation rules and sequences, then generates a plan document directly usable by coding agent, with iterative optimization based on feedback from subsequent stages.

COPA bridges detailed requirements/architecture documents and actual coding. Its core role is to convert SRS and ADD content into structured, actionable code implementation plans to guide subsequent coding. Inputs are detailed SRS and ADD; output is a specific code implementation plan for coding and a structured project framework. To achieve this, COPA draws on the \textbf{SRS \& ADD Knowledge} base, which provides \textbf{SRSs} (functional requirements, user stories, acceptance criteria), \textbf{ADDs} (module decomposition, interface contracts, design patterns, technology stack), and \textbf{Standards} (IEEE 830~\cite{DBLP:conf/worldcist/ChikhA14}, ISO 29148~\cite{DBLP:journals/csi/GarciaPLC20} templates and traceability matrices). Leveraging these resources, COPA performs a thorough analysis, extracts key information, transforms it into specific code implementation rules and sequences, and generates a plan document directly usable by the coding agent, with iterative optimization based on feedback from subsequent stages.

\textbf{SRS \& ADD Analysis:}
COPA first thoroughly analyzes detailed SRS and ADD, focusing on converting business processes into code implementation plans. It extracts specific processes and constraints from the SRS. From the ADD, it retrieves details of module division, analyzes module interface specifications and related technical constraints, and technology stack details. This analysis will also identify ambiguities, ensuring a consistent understanding of requirements before planning. Subsequently, it evaluates the positioning of these extracted elements in actual coding and prepares a code implementation plan before coding.

\textbf{Generation of Code Implementation Planing (Code Plan):}
Based on the analysis results, COPA will generate a detailed code implementation plan document specifically designed for coding. This document not only clarifies the order of code implementation, but also specifies the arrangement rules for methods within each module, such as class inheritance hierarchy, access modifiers for methods and properties. This will serve as a direct operational manual for the subsequent coding agent. It provides the component implementation sequence and module integration logic for coding. The structured design of the document facilitates modifications, allowing for optimization based on feedback.

\textbf{Project Structure Generation:} COPA will transform the implementation plan into specific project structures. This includes defining package levels and outlining directory layouts for source code, test files, and resources. It also predefines module entry points and dependency management configurations, ensuring that coding agents can directly populate content into well-structured frameworks. This structure aligns with the implementation plan and industry best practices to ensure maintainability.
\vspace{-0.5em} 
\subsection{Coding Agent (CA)}

CA is responsible for generating executable source code. Its core role is to produce code that aligns with the plan, adheres to specific coding standards and style guidelines, and is optimized based on feedback. CA's input includes the code implementation plan and project structure provided by COPA; its output is source code that meets the requirements, along with an API list of third-party libraries used. To accomplish this, CA first extracts the dependency context surrounding the target fragment, queries the Coding Knowledge base, analyzes required third-party APIs, generates code according to the input plan while following coding standards, then compiles and debugs the code, and finally iteratively optimizes the code based on feedback.

CA consults a dedicated knowledge base composed of five pillars:
\textbf{Open Source Projects} (curated GitHub/GitLab~\cite{DBLP:journals/ploscb/BlischakDW16} repositories classified by architecture style and domain),
\textbf{API Library} (third-party libraries with usage patterns, version constraints, and compatibility notes),
\textbf{Domain Experts} (heuristics extracted from maintainer annotations, Stack Overflow discussions, and official documentation),
\textbf{Coding Standards} (language-specific conventions, secure coding guidelines, and design patterns),
and \textbf{Coding Tools} (static analyzers, linters, template engines, and build automation tools).
This corpus ensures that generated code is syntactically correct, idiomatic, secure, and reusable while remaining fully aligned with the COPA plan.

\textbf{Third-Party API Analysis:}
Before code generation, CA analyzes the technical specifications and functional requirements listed in COPA's code implementation plan to determine the necessary third-party API libraries. Compile a list of APIs, detailing the library names, version constraints, and specific modules/classes to be used, ensuring compatibility with the project's architecture and coding standards. This list can facilitate tracing code issues.

\textbf{Code Generation:}
Receiving the technical specifications and CA follows the arrangement, gradually builds the project structure, organizes file order, carefully implements code, and ensures strict adherence to coding standards and technical frameworks. CA maintains consistency with COPA to generate all necessary code files. This process utilizes predefined code templates and design pattern libraries to ensure architectural consistency, and all generation steps closely reference detailed information in the COPA plan.

\textbf{Code Compilation and Self-Debugging:}
Post-generation, CA employs just-in-time compilation to validate syntax correctness and identify basic issues in new files. It records detailed compilation logs, including error types, locations, and suggested fixes, to facilitate troubleshooting. Upon successful compilation of each file, CA proceeds to the next in the sequence defined by COPA’s dependency graph. After generating all codes and configuration files, CA uses appropriate build tools to compile and execute the application. Verifies the fundamental launch capability and operational state, acting as a ``self-check'' for basic functionality, such as ensuring that core modules initialize correctly.

\textbf{Code Rectification:}
After receiving a compilation error report, TA's test report, or its own code quality inspection report, CA initiates a comprehensive code optimization process to resolve errors. First, it analyzes the code implementation plan, error context, and API list to determine the root cause. It then generates targeted code fix snippets based on feedback suggestions and, if necessary, adjusts API usage. It maintains consistent coding style, recompiles to verify the fixes, and ensures optimizations do not introduce new issues. Throughout the process, CA ensures that the corrected code remains aligned with the COPA plan, updates the code quality report, compilation logs, and API list to reflect optimization measures, and creates traceable improvement records.

% \ws{The testing agent should clearly define the test target, testing methodology, test generation (including inputs and oracles), test evaluation, and test reporting.}
\vspace{-0.5em} 
\subsection{Testing Agent (TA)}
% TA is responsible for verifying the quality of code generated by CA through unit testing. The core responsibilities include using SRS and ADD to analyze test boundaries and scope, conducting targeted unit tests, and generating test reports. Inputs include the code generated by CA, as well as SRS and ADD. The output is a test report.
TA is responsible for verifying the quality of code generated by CA through unit testing. The core responsibilities include using SRS and ADD to analyze test boundaries and scope, conducting targeted unit tests, and generating test reports. Inputs include the code generated by CA, as well as SRS and ADD. The output is a test report.

TA consults a dedicated knowledge base that contains four pillars:
\textbf{Testing Projects} (historical test suites and plans reused for regression benchmarking and coverage comparison),
\textbf{Testing Criteria} (requirement-to-test mappings, coverage objectives, and defect classification schemas per IEEE 829~\cite{DBLP:conf/icsecs/SidekNW11} and ISTQB~\cite{DBLP:books/sp/Roman18}),
\textbf{Testing Standards} (framework-specific best practices, test design techniques, and assertion patterns),
and \textbf{Testing Tools} (automated test generators, mutation testing frameworks, coverage analyzers, and continuous integration hooks).
This corpus enables TA to derive meaningful test cases, detect faults early, and produce structured, traceable feedback that drives iterative code refinement.

% \textbf{Test Boundary and Scope Analysis:}
% TA first conducted a detailed analysis of SRS and ADD to clearly define the boundaries and scope of unit testing. The test cases were derived from user stories in the SRS and CA's source code. It determines the test inputs and expected test oracles and selects an appropriate testing framework. TA mapped the functional requirements of the SRS to specific code modules, identifying the core logic, module boundaries, component responsibilities, and technical specifications for each method.

% \textbf{Unit Testing:} 
% TA uses an appropriate testing framework to execute unit tests. It verifies whether the behavior of each method and class complies with the specifications: checking functional correctness, adherence to interface standards, and handling of edge cases. TA records detailed test results, including pass/fail status, error messages, and stack traces of failed cases.

\textbf{Test Plan Generation:}
TA first conducts a detailed analysis of SRS and ADD to clearly define the boundaries and scope of testing. The test plan is derived from user stories in the SRS and CA's source code. It determines the test inputs and expected test oracles, and selects an appropriate testing framework. TA maps the functional requirements of the SRS to specific code modules, identifying the core logic, module boundaries, component responsibilities, and technical specifications for each method.

\textbf{Test Case Generation:}
TA first parses both the SRS user stories and the ADD component contracts to identify equivalence classes, boundary values, and exception-triggering conditions for each method. It then produces concrete JUnit~\cite{DBLP:books/daglib/0044051} (or PyTest~\cite{okken2022python}, etc.) test methods: one positive case per nominal path, one negative case per invalid input class, and one boundary pair per numeric limit. Templates enriched with guard-assertions and mocked dependencies are instantiated automatically; random or property-based tests are added when the specification is nondeterministic. All generated cases are annotated with traceability links (requirement ID + method signature) so that any subsequent change in SRS/ADD prompts automatic regeneration of the affected tests.

\textbf{Test Evaluation and Reporting:}
After the test execution, TA evaluates the unit test results to identify code defects. It compiles a test report that not only does it include test coverage, pass/fail status, and identified defects, but it also ensures compliance with industry testing standards, clearly records test objectives, maintains consistent defect classification criteria, and standardizes issue severity levels. These reports maintain traceability with SRS requirements and ADD component definitions, making them easy to review. TA also dynamically manages test cases, scripts, and reports, updating them as changes occur in the SRS, ADD, or generated code.
\vspace{1em}

%% file: sections/case_study.tex
\vspace{-0.5em}

\section{Case Study}

\label{sec:case_study}
% \ws{Keep the Case Study content within half a page (i.e., one column).}

\begin{figure*}[!t]
    \centering
    \includegraphics[width=0.79\textwidth]{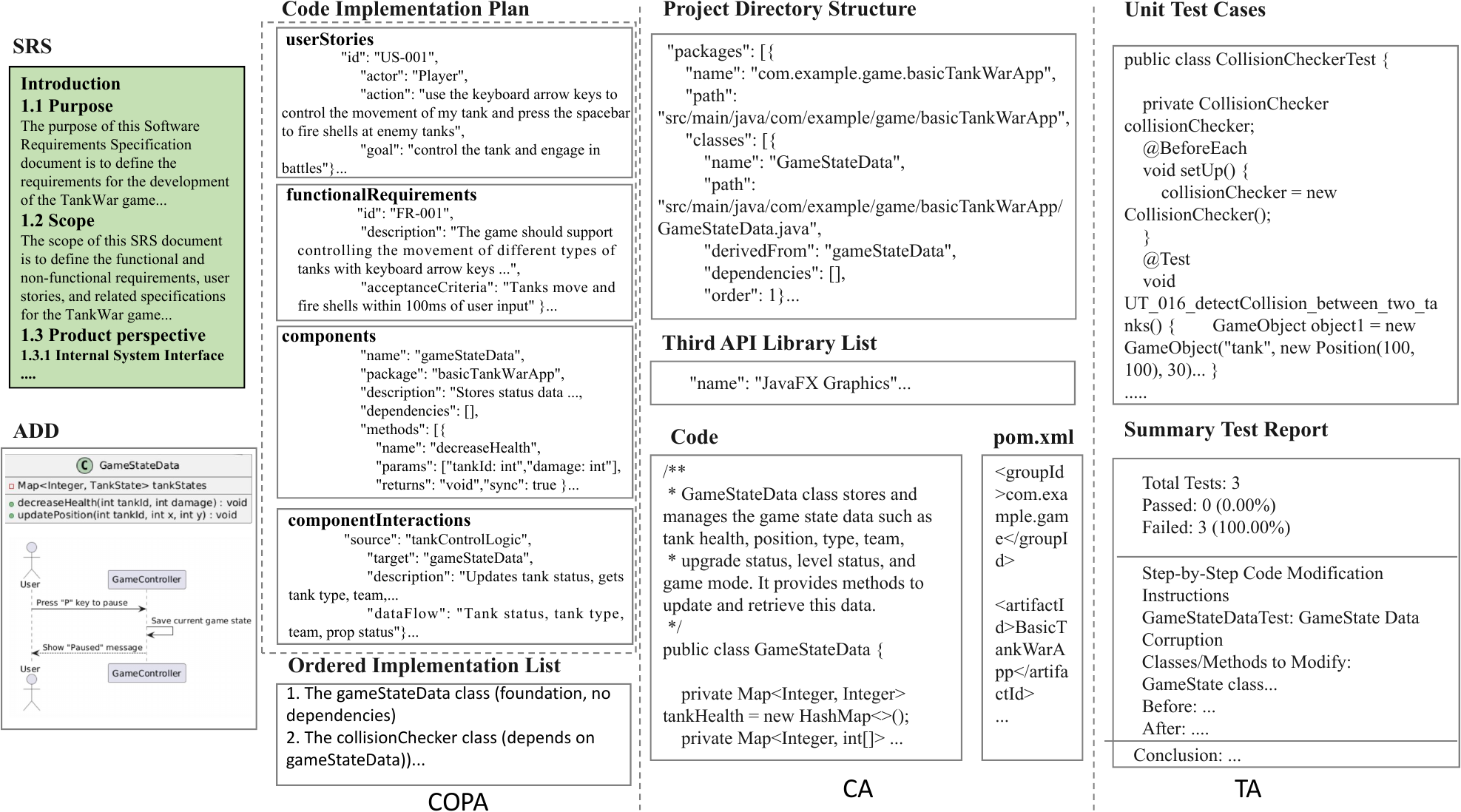}
    \caption{An Example of Proposed Code Generation Process}
    \label{F:Case}
\end{figure*}

To facilitate understanding, Figure~\ref{F:Case} illustrates an example of the development of a Java Tank Battle game. In this example, the COPA first analyzes the input SRS(e.g., project goals and scenarios for user stories) and ADD(e.g., design patterns, strategies, and logical views), extracting key elements such as core gameplay (e.g., players controlling tanks to shoot enemies), technical constraints (e.g., selection of JavaFX as the graphics framework) and module responsibilities. COPA defines method interface specifications (e.g., input/output rules for the \texttt{decreaseHealth(int tankId, int damage)} method) and establishes a logical code implementation order aligned with module dependencies. COPA then outputs plan documents in JSON format, including component dependency graphs, hierarchical package structures (e.g., the core package \texttt{basicTankWarApp} with subpackages like \texttt{.logic}, \texttt{.model} and \texttt{.view} for functional partitioning), and version control markers.

Next, the CA generates specific code frameworks based on COPA’s artifacts. For instance, it creates the ``GameStateData'' class under the ``logic'' subpackage, defining variables like \texttt{tankHealth} and \texttt{tankPosition}, and integrates third-party APIs (e.g., JavaFX Graphics). During the compilation verification phase, CA resolves issues such as uninitialized instances (e.g., NullPointerException in ``CollisionChecker'' due to uninitialized \texttt{tankPosition} in ``GameStateData'') through a feedback loop, either by requesting COPA to update code implementation plans or self-debugging.  

TA then clarifies the scope of the testing, selecting JUnit as the testing framework and designing unit tests (e.g., \texttt{GameStateDataTest} and \texttt{CollisionCheckerTest}) based on the COPA plans and CA's source code. Using equivalence partitioning and boundary value analysis, TA generates test cases such as verifying \texttt{decreaseHealth} with damage values at the tank health boundary (e.g., damage exceeding remaining health). For example, TA identifies a defect in the \texttt{checkCollision} method’s coordinate calculation logic, which fails to detect diagonal tank collisions, and provides a structured test report detailing the issue, including the test input, expected oracle, actual result. As COPA modifies requirements, TA dynamically adjusts test cases to ensure coverage of the updated logic.

%% file: sections/challenges_and_opportunities.tex
\section{Challenges and Opportunities}
\label{sec:challenges_and_opportunities}
Grounded in the \ours{} framework, we outline these issues and opportunities as follows.
\textbf{For COPA:}
A primary challenge is the precise translation of SRS and ADD into actionable code implementation plans while avoiding erroneous relationship mappings. Corresponding opportunities include the development of refined implementation plans that enhance code generation, standardize the generation sequence, and minimize errors.
\textbf{For CA:}
Key challenges encompass strict adherence to standards and plans during generation, effective refactoring of code in response to COPA updates without introducing new defects, and timely remediation of flaws identified by TA testing. Opportunities lie in boosting efficiency through predefined templates and third party API libraries, as well as in leveraging feedback for continuous code optimization.
\textbf{For TA:}
The central challenge is the design of comprehensive unit tests that cover SRS scenarios, coupled with the dynamic updating of test cases while ensuring their quality whenever COPA plans evolve. The attendant opportunity is that rigorous testing can elevate code generation quality and curtail the time spent on blind debugging.
\textbf{For \ours{}:}
Challenges include securing efficient inter agent coordination and seamless communication, together with the capacity to adapt to evolving requirements and to inject pertinent knowledge that supports agent tasks. Opportunities reside in exploiting the interpretive power of LLMs through well orchestrated coordination protocols, thereby enabling effective information exchange and feedback integration and markedly improving code generation quality. Although each component of the \ours{} framework confronts distinct obstacles, corresponding opportunities exist to overcome them, positioning the framework to advance code generation quality via coordinated effort and the strategic deployment of cutting edge techniques.

%% file: sections/conclusion.tex
\section{Conclusion}
\label{Sec:Conclusion}
% \ws{The conclusion section should be concise and limited to approximately ten lines.}
This paper introduces \ours{} framework, which employs three specialized agents, i.e., COPA, CA, and TA, to transform SRS and ADD into executable code through iterative feedback. COPA generates implementation plans and project structures, CA develops code using these plans and feedback for debugging, and TA defines test boundaries, executes unit tests, and reports to ensure quality. By leveraging LLMs and coordinating agent collaboration, the framework advances towards full automation in application-level software development.